\documentstyle[sprocl,epsf,rotate]{article}

\input{psfig}

\def\Journal#1#2#3#4{{#1} {\bf #2}, #3 (#4)}

\def\PRL{\em Phys. Rev. Lett.}

\def\PRC{{\em Phys. Rev.} C}

\def\EPA{{\em Eur. Phys. J.} A}
\def\RPP{\em Rep. Prog. Phys.}

\def\be{\begin{equation}}
\def\ee{\end{equation}}
\def\bea{\begin{eqnarray}}
\def\eea{\end{eqnarray}}

\newcommand{\hbo}{\hbar \omega}

\begin{document}

\title{PARTIAL DYNAMICAL SYMMETRIES IN NUCLEI}

\author{A. LEVIATAN}

\address{Racah Institute of Physics, The Hebrew University, Jerusalem 91904, 
Israel}


\maketitle\abstracts{
Partial dynamical symmetries (PDS) are shown to be relevant to the 
interpretation of the $K=0_2$ band and to the occurrence of F-spin 
multiplets of ground and scissors bands in deformed nuclei. 
Hamiltonians with bosonic and fermionic PDS are presented.}

\section{Introduction}
When a dynamical symmetry occurs all properties of the system (e.g. energy
eigenvalues) and wave functions are known analytically thus providing 
clarifying insights into complex dynamics. 
The majority of nuclei, however, do not satisfy the predictions of exact 
dynamical symmetries. Instead, one often finds that 
only a subset of states fulfill the symmetry while other states do not. 
In such circumstances, referred to as partial symmetries, 
the Hamiltonian supports a coexistence of ``special'' solvable 
states and other states which are mixed. Examples of partial symmetries 
in nuclear spectra are discussed below. 

\section{Partial SU(3) symmetry and the nature of the $K=0_2$ band}
The nature of the lowest K=$0^{+}$ [K=$0_{2}$] excitation in deformed 
nuclei is still subject to controversy. Traditionally described as 
a $\beta$ vibration, its properties  
are empirically erratic in contrast to the 
regular behavior observed for ground and $\gamma$ bands.
This suggests different symmetry character for these bands. With that 
in mind, the following IBM \cite{ibm87} Hamiltonian with partial 
SU(3) symmetry has been proposed \cite{lev96}
\bea
H \;=\; h_{0}P^{\dagger}_{0}P_{0} + h_{2}P^{\dagger}_{2}
\cdot\tilde P_{2}~, 
\label{hpds}
\eea
where $P^{\dagger}_{L}$ $(L=0,2)$ are boson-pairs. 
Although $H$ is not an $SU(3)$ scalar, it has solvable ground and $\gamma$ 
bands with good $SU(3)$ symmetry, $(\lambda,\mu)=(2N,0),\,(2N-4,2)$ 
respectively. In contrast, the $K=0_2$ band involves a mixture of 
$SU(3)$ irreps $(2N-4,2)$, $(2N-8,4)$ and $(2N-6,0)$ or equivalently a 
mixture of single-phonon ($\beta$) and double-phonon ($\gamma^2_{K=0}$ 
and $\beta^2$) components. The respective probability amplitudes 
$(A_{\beta})^2$, $(A_{\gamma^2})^2$, $(A_{\beta^2})^2$ are shown in 
Fig.~1.
\begin{figure}[t]
\rotate{\rotate{\rotate{
\psfig{figure=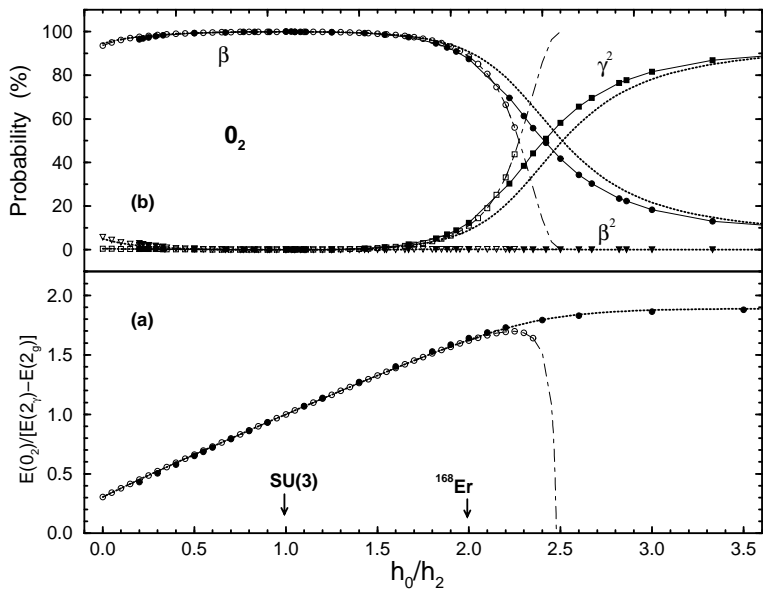,width=0.5\linewidth}}}}
\vspace{3.0truecm}
\caption{Properties of the $K=0_2$ band as a function of $h_0/h_2$, 
parameters of the SU(3) PDS Hamiltonian, Eq.~(\ref{hpds}), N=16 (solid lines).
The dotted and dot-dashed lines are approximations based on 3-band mixing 
calculation. (a) Ratio of $K=0_2$ and $\gamma$ bandhead energies. 
(b)~Probability amplitudes squared, $(A_{\beta})^2$, $(A_{\gamma^2})^2$, 
$(A_{\beta^2})^2$ for the $K=0_2$ wave function. Figure taken from [3].
\label{fig1}} 
\end{figure}
In the current PDS scheme both  
the SU(3) breaking and the double-phonon admixture in the 
$K=0_2$ wave function are given by $(1-A_{\beta}^2)$. 
The mixing is of order ($1/N)$ but 
depends critically on the ratio of the $K=0_2$ and $\gamma$ bandhead 
energies (also shown in Fig. 1). For most of the relevant range of 
$h_0/h_2$, corresponding to bandhead ratio in the range $0.8-1.65$, 
the double-phonon admixture is at most $\sim 15\%$ ($12.5\%$ in $^{168}$Er).
These findings support the conventional single-phonon 
interpretation for the $K=0_2$ band with small but significant 
double-$\gamma$-phonon admixture. 
\begin{table}[t]
\caption{Comparison of theoretical and experimental absolute B(E2) 
values [W.u.] for transitions from the $2^{+}_{K=0_2}$ level [4] and  
to the $0^{+}_{K=0_2}$ level [5] in $^{168}$Er.\label{tab1}}
\vspace{0.2cm}
\begin{center}
\footnotesize
\begin{tabular}{|lcc|ccc|}
\hline
\multicolumn{3}{|c|}{Exp.} &
\multicolumn{3}{c|}{Calc.}\\
Transition & B(E2) & range & PDS & WCD & CQF \\
\hline
\multicolumn{3}{|c|}{Lifetime measurement [4]} &
\multicolumn{3}{c|}{ }\\
$2^{+}_{K=0_2}\rightarrow 0^{+}_g$ & 0.4 & 0.06--0.94  &
0.65  & 0.15  & 0.03  \\
$2^{+}_{K=0_2}\rightarrow 2^{+}_g$ & 0.5 & 0.07--1.27  &
1.02  & 0.24  & 0.03  \\
$2^{+}_{K=0_2}\rightarrow 4^{+}_g$ & 2.2 & 0.4--5.1  &
2.27 & 0.50 & 0.10 \\
$2^{+}_{K=0_2}\rightarrow 2^{+}_{\gamma}$ $^{a)}$ & 6.2 (3.1) 
& 1--15 (0.5--7.5) &
4.08 & 4.16 & 4.53 \\
$2^{+}_{K=0_2}\rightarrow 3^{+}_{\gamma}$ $^{a)}$ & 7.2 (3.6) 
& 1--19 (0.5--9.5) &
7.52 & 7.90 & 12.64 \\
\hline
\multicolumn{3}{|c|}{Coulomb excitation [5]} &
\multicolumn{3}{c|}{ }\\
$2^{+}_g\rightarrow 0^{+}_{K=0_2}$ & $0.08\pm 0.01$  & &
0.79   & 0.18   & 0.03  \\
$2^{+}_{\gamma}\rightarrow 0^{+}_{K=0_2}$ & $0.55\pm 0.08$  & &
3.06 & 3.20 & 5.29 \\
\hline
\end{tabular}
\end{center}
\end{table}
\noindent
Since the wave functions of the solvable states are known, it is
possible to obtain analytic expressions for the E2 rates
between them \cite{lev96,levsin99}. 
B(E2) ratios for $\gamma\rightarrow g$ transitions are parameter-free 
predictions of SU(3) PDS, and have been used \cite{lev96} to establish 
the validity of this scheme in $^{168}$Er. Absolute B(E2) values for 
$K=0_2\rightarrow g$ transitions can be used \cite{levsin99} to 
extract $(A_{\beta})^2$.
In Table 1 we compare the predictions of the PDS and broken-SU(3) 
calculations: added $O(6)$ term (WCD) and consistent-Q formalism (CQF), 
with the B(E2) values 
deduced from a lifetime measurement~\cite{lehmann98} and Coulomb 
excitation \cite{hart98} in $^{168}$Er.
It is seen that the PDS 
and WCD calculations agree well with the lifetime measurement, but 
the CQF calculation under-predicts the 
$K=0_2\rightarrow g$ data. 
On the other hand, all calculations show large deviations from the 
quoted B(E2) values measured in Coulomb excitation. It should be noted, 
however, that there are serious discrepancies between the above two 
measurements \cite{levsin99}. An independent measurement of the lifetime 
of the $0^{+}_{K=0_2}$ in $^{168}$Er is highly desirable to clarify 
this issue.

\section{F-spin as a partial symmetry}
F-spin characterizes the proton-neutron ($\pi$-$\nu$) symmetry of 
IBM-2 states. There are empirical indications \cite{lipas90} 
that low lying collective states have predominantly 
$F=F_{max}=(N_{\pi}+N_{\nu})/2$ with typical impurities of $2\%-4\%$. 
In spite of its appeal, however, F-spin cannot be an exact symmetry of the 
Hamiltonian. The assumption of F-spin scalar Hamiltonians is at variance 
with the microscopic interpretation of the IBM-2, 
which necessitates different effective 
interactions between like and unlike nucleons. 
Furthermore, if F-spin was a symmetry of the Hamiltonian, 
then {\it all} states 
would have good F-spin and would be arranged in F-spin multiplets. 
Experimentally the latter are observed in ground bands but not necessarily 
in excited  $\beta$ and $\gamma$ bands.
Thus F-spin can at best be an approximate quantum number which is 
good only for a selected set of states.
These are precisely the signatures of a partial symmetry.
\begin{table}[t]
\caption{
The ratio $R=\sum B(M1)\uparrow/(C_{F,F_0})^2$ 
for members of F-spin~multiplets. Here 
$C_{F,F_0} =~(F,F_0;1,0\vert F-1,F_0)$. 
The low value of the summed M1 strength to the scissors mode 
$\sum B(M1)\uparrow $ 
for $^{172}$Yb has been attributed to experimental deficiencies.} 
\vspace{0.2cm}
\begin{center}
\footnotesize
\begin{tabular}{|lccccc|}
\hline
Nucleus & $F$ & $F_0$ & $\sum B(M1)\uparrow$ $[\mu_{N}^2]$ 
& $(C_{F,F_0})^2$ & $R$ \\
\hline
$^{148}$Nd & 4    & 1    & 0.78 (0.07) & 5/12     & 1.87 (0.17) \\
$^{148}$Sm &      & 2    & 0.43 (0.12) & 1/3      & 1.29 (0.36) \\
\hline
$^{150}$Nd & 9/2  & 1/2  & 1.61 (0.09) & 4/9      & 3.62 (0.20) \\
$^{150}$Sm &      & 3/2  & 0.92 (0.06) & 2/5      & 2.30 (0.15) \\
\hline
$^{154}$Sm & 11/2 & 1/2  & 2.18 (0.12) & 5/11     & 4.80 (0.26) \\
$^{154}$Gd &      & 3/2  & 2.60 (0.50) & 14/33    & 6.13 (1.18) \\
\hline
$^{160}$Gd & 7    & 0    & 2.97 (0.12) & 7/15     & 6.36 (0.26) \\
$^{160}$Dy &      & 1    & 2.42 (0.18) & 16/35    & 5.29 (0.39) \\
\hline
$^{162}$Dy & 15/2 & 1/2  & 2.49 (0.13) & 7/15     & 5.34 (0.28) \\
$^{166}$Er &      & $-1/2$ & 2.67 (0.19) & 7/15   & 5.72 (0.41) \\
\hline
$^{164}$Dy & 8    & 0    & 3.18 (0.15) & 8/17     & 6.76 (0.32) \\
$^{168}$Er &      & $-1$   & 3.30 (0.12) & 63/136 & 7.12 (0.26) \\
$^{172}$Yb &      & $-2$   & 1.94 (0.22) & 15/34 & 4.40 (0.50) \\
\hline
$^{170}$Er & 17/2 & $-3/2$ & 2.63 (0.16) & 70/153 & 5.75 (0.35) \\
$^{174}$Yb &      & $-5/2$ & 2.70 (0.31) & 66/153 & 6.26 (0.72) \\
\hline
\end{tabular}
\end{center}
\end{table}
A class of IBM-2 Hamiltonians with such property 
has been proposed \cite{levgin00}
\bea
H &=& 
\sum_{i}\sum_{L=0,2}
A^{(L)}_{i}R^{\dagger}_{i,L}\cdot\tilde{R}_{i,L} 
+ B\hat{{\cal M}}_{\pi\nu}
\label{hprime}
\eea
The $R^{\dagger}_{i,L}$ ($L=0,2$) 
are boson pairs with $F=1,\,i=F_0=0,\pm 1$
and $\hat{{\cal M}}_{\pi\nu}$ is the Majorana operator.  
The above Hamiltonian is non-F-scalar but has 
a subset of {\it solvable} states which form F-spin multiplets for 
the $K=0$ ground 
band with $F=F_{max}$, and for the $K=1$ scissors band 
with $F=F_{max}-1$, while other excited bands are mixed.
For ground bands such structures have 
been empirically established. 
Since the $M1$ operator $(\hat{L}_{\pi}- \hat{L}_{\nu})$ 
is an F-spin vector, the prediction for F-spin multiplets of scissors 
states can be tested by examining the ratio of summed ground to scissors 
$B(M1)$ strength divided by the square of the appropriate 
Clebsch Gordan coefficient. In Table 2 we list 
{\it all} F-spin partners for which $\sum B(M1)\uparrow$ 
has been measured todate. 
It is seen that within the experimental errors, the above ratio is, as 
expected, fairly constant. The solvable ground and scissors bands have 
the same moment of inertia in agreement with the conclusions of a recent 
comprehensive analysis of the scissors mode in heavy even-even 
nuclei \cite{enders99}.

\section{Fermionic Partial Symmetry}

Partial symmetries are not confined to bosonic systems. 
A fermionic Hamiltonian with SU(3) partial symmetry has been proposed 
\cite{escherlev00} in the framework of the symplectic shell model 
\cite{sympl}, 
\bea
H(\beta_0,\beta_2) = \beta_0 \hat{A}_0 \hat{B}_0
+ \beta_2 \hat{A}_2 \cdot \hat{B}_2 ~,
\label{Eq:Hpds} 
\eea
with a structure similar to that of the bosonic Hamiltonian of 
Eq. (\ref{hpds}). The $\hat{A}_{L}$ ($\hat{B}_{L}$), $L$ = 0 or 2,
are symplectic generators which create (annihilate) $2 \hbo$ excitations 
in the system. The above Hamiltonian is not SU(3) invariant 
but has a subset of solvable pure-SU(3) states ({\it e.g.} the 
$0\hbo$ $K=0_1$ and $2\hbo$ $K=2_1$ bands in Fig.~2). The PDS 
Hamiltonian (\ref{Eq:Hpds}) can be rewritten in terms of the symplectic 
quadrupole-quadrupole interaction $Q_{2}\cdot Q_{2}$ plus terms diagonal 
in the Sp(6,R) $\supset$ SU(3) $\supset$ SO(3) chain and terms coupling 
different harmonic oscillator shells. The eigenstates of the two 
Hamiltonians are 
compared in Fig. 2 with parameters tuned to the ground band of $^{20}$Ne.
For both the ground and the resonance bands, PDS eigenstates are seen
to approximately reproduce the structure of the exact $Q_2 \cdot Q_2$
eigenstates within the $0 \hbo$ and $2 \hbo$ spaces, respectively.
In particular, for each pure state of the PDS scheme we find a
corresponding eigenstate of the quadrupole-quadrupole interaction,
which is dominated by the same SU(3) irrep.
Moreover, for reasonable interaction parameters, each rotational band
is primarily located in one level of excitation, with the exception of
the lowest $K=0_2$ resonance band, which is spread over many
$N\hbo$ excitations.
\begin{figure}[t]
\psfig{figure=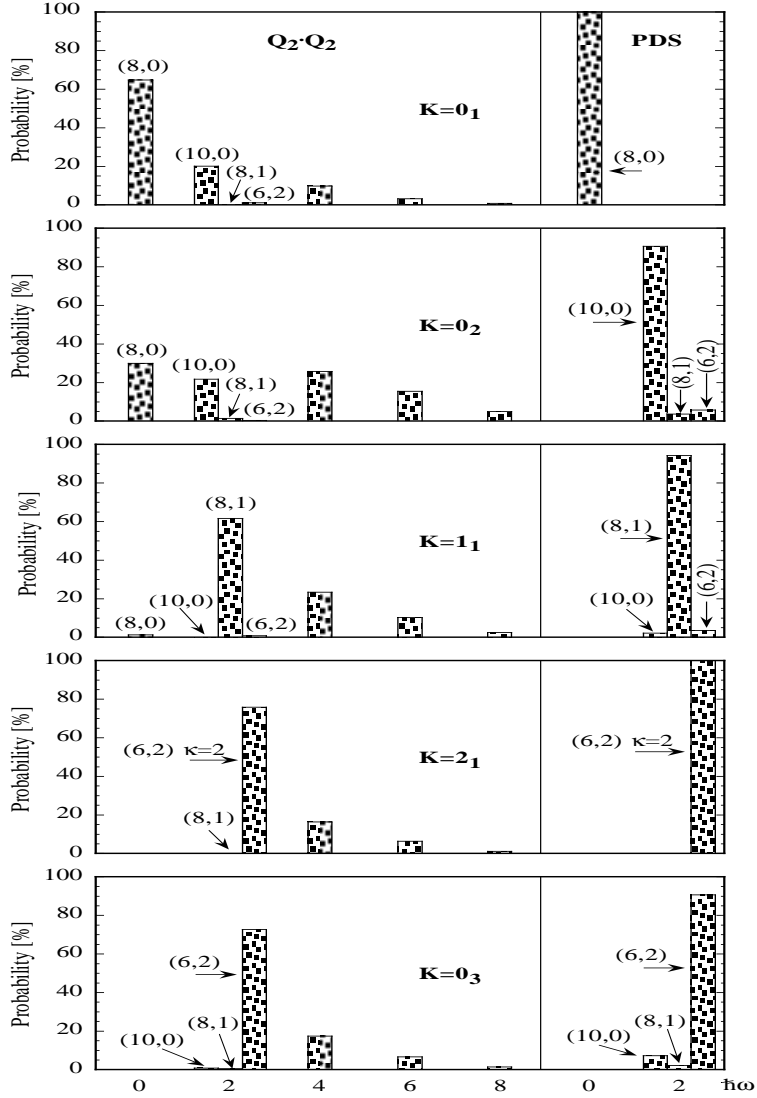,height=15.5truecm,width=1.5\linewidth}
\caption{Decomposition for calculated $2^+$ states of $^{20}$Ne.
Individual contributions from the relevant SU(3) irreps at the 0$\hbo$
and 2$\hbo$ levels are shown for both a symplectic $8\hbo$ calculation
(denoted $Q_{2}\cdot Q_{2}$) and a PDS calculation.
In addition, the total strengths contributed by the $N\hbo$ excitations
for $N>2$ are given for the symplectic case. Figure taken from [9].
\label{Decomp}}
\end{figure}

\section*{Acknowledgments}
This research was supported by a grant from the United States-Israel 
Binational Science Foundation (BSF), Jerusalem, Israel. The works reported 
in Sections 2,\,3,\,4 were done in collaboration with I. Sinai (HU), J.N. 
Ginocchio (LANL) and J. Escher (HU,TRIUMF) respectively.

\end{document}